\documentclass[a4paper,12pt]{article}

\usepackage{amsmath}
\usepackage{afterpage}
\usepackage{booktabs}
\usepackage{bm}
\usepackage{bbm}
\usepackage[justification=justified]{caption}
\usepackage{float}
\usepackage{fullpage}
\usepackage{geometry}
\usepackage[bookmarks, hidelinks]{hyperref}
\usepackage{xurl}
\usepackage[round]{natbib}
\usepackage{pdflscape}
\usepackage{setspace}
\usepackage{subcaption}
\usepackage{tabularx}
\usepackage[textsize=footnotesize]{todonotes}
\usepackage{xcolor}
\usepackage{comment}

\usepackage[english]{babel}
\usepackage{amsthm}
\usepackage{amssymb}
\usepackage{mathtools}

\DeclareMathOperator{\arsinh}{arsinh}

\newtheorem{proposition}{Proposition}

\newcommand{\beginappendixA}{%
        \setcounter{table}{0}
        \renewcommand{\thetable}{A\arabic{table}}%
        \setcounter{figure}{0}
        \renewcommand{\thefigure}{A\arabic{figure}}%
     }
\newcommand{\beginappendixB}{%
        \setcounter{table}{0}
        \renewcommand{\thetable}{B\arabic{table}}%
        \setcounter{figure}{0}
        \renewcommand{\thefigure}{B\arabic{figure}}%
     }
\newcommand{\beginappendixC}{%
        \setcounter{table}{0}
        \renewcommand{\thetable}{C\arabic{table}}%
        \setcounter{figure}{0}
        \renewcommand{\thefigure}{C\arabic{figure}}%
     }
\newcommand{\beginappendixD}{%
       \setcounter{table}{0}
       \renewcommand{\thetable}{D\arabic{table}}%
       \setcounter{figure}{0}
       \renewcommand{\thefigure}{D\arabic{figure}}%
    }

\newcolumntype{b}{>{\hsize=2\hsize}X}
\newcolumntype{m}{>{\hsize=1.6\hsize}X}
\newcolumntype{s}{>{\hsize=.6\hsize}X}

\providecommand{\keywords}[1]
{
  \small
  \textbf{\textit{Keywords---}} #1
}

\providecommand{\jelcodes}[1]
{
  \small
  \textbf{\textit{JEL Codes---}} #1
}

\graphicspath{{../graphs/}}

\title{What's Logs Got to do With it: On the Perils of log Dependent Variables and Difference-in-Differences\thanks{I benefited a great deal from discussions about this work with Jaime Mill\'an-Quijano. I am also very grateful to Marc F. Bellemare for detailed comments on an early draft. McConnell: Southampton, \href{mailto: brendon.mcconnell@gmail.com}{\texttt{brendon.mcconnell@gmail.com}}.
  }}

\author{Brendon McConnell}
\date{\today}

\begin{document}
\maketitle

\begin{abstract}
The logarithmic transformation of the dependent variable is not innocuous when using a difference-in-differences (DD) research design. 
With a dependent variable in logs, the DD term does not capture the outcome difference between treated and untreated groups over time.
Rather it reflects an approximation of the proportional difference in growth rates across groups. 
As I show with both simulations and two empirical examples, if the baseline outcome distributions are sufficiently different across groups, the DD parameter for a log-specification can be different in sign to that of a levels-specification. I provide a condition, based on (i) the aggregate time effect, and (ii) the difference in relative baseline outcome means, for when the sign-switch will occur. \end{abstract}
\vfill
 \\
\jelcodes{C01.}

\clearpage
\section{\label{sec:Intro}Introduction}
Difference-in-differences (DD) is almost certainly the most popular quasi-experimental research design currently used in a broad range of empirical settings. Its use extends beyond Economics into Political Science, Social Medicine and other fields. The seeming simplicity of the design has, at least until recently, played a large role in its popularity. A recent literature  documenting  underlying issues with difference-in-differences, particularly in the case of staggered roll-out of treatment implementation \citep{GoodmanBacon2021}, has somewhat shattered the illusion of the simplicity of this research design.

In this paper, I highlight an additional complication that the applied researcher faces when operationalizing a DD design -- the choice of functional form, and the subsequent consequences of this choice. As \citet{Ciani2019} note in related work, researchers may apply the log transformation, not because they believe the true underlying data generating process is multiplicative rather than additive, but rather due to concerns of skewness, or because the log transformation enables one to interpret the effect of controls in percentage terms.\footnote{Some papers are more explicit about the consequences of the choice between level- and log-specifications of the outcome variable when using difference-in-differences designs \citep{Finkelstein2007,Powell2020,Park2021}.
}

In order to set the scene for this work, I survey all papers published in \emph{The Quarterly Journal of Economics (QJE)} from 2018 to 2022. I present summary information on these articles in Table~\ref{tab:QJE_2018_2022_1}. Of the 49 \emph{QJE} articles published in this five year span that use a DD design, almost all of these (46 articles) consider at least one continuous outcome variable.  Just under two thirds of these 46 articles (30 articles or 65\%) impose a log transformation on at least one continuous variable, and three more impose an inverse hyperbolic sine transformation\footnote{See \citet{Bellemare2020} for an in-depth consideration of the implications of the inverse hyperbolic sine (IHS) transformation. As the authors note, a key driver of the recent interest from applied researchers in the IHS transformation is that the function is similar to a logarithm, but unlike the logarithmic function, it is defined at zero.}, which will lead to related issues.\footnote{The inverse hyperbolic sine of a random variable $y$, $\arsinh(y) = \ln(y +\sqrt{y^2+1}) \approx \ln(2y)$ for large $y$.} That over 60\% of all DD papers in one of the top Economics journals use at least one log-dependent variable specification underscores the importance of better understanding what we recover using a DD design with a log transformed dependent variable.

I start by outlining the key differences between a DD model with level- and log-dependent variables. In the level-dependent variables case, the DD parameter returns the difference  between the treated and untreated groups in changes in the outcome variable over time. In contrast, when one uses a log transformation of the dependent variable, the DD parameter approximates the  \textit{proportional difference in growth rates} between treated and untreated groups.\footnote{Denoting the DD parameter for the log specification as $\beta_4$, the transformation $\exp(\beta_4)-1$ yields the precise proportional difference in growth rates.} This difference in the two specifications can yield highly disparate DD estimates when the distributions of the treated and untreated groups are sufficiently different in the pre-policy period.\footnote{\citet{Meyer1995} warned of using DD specifications in cases where the distribution of outcomes for treated and untreated groups was sufficiently different in the pre-policy period, noting in particular the issue of non-linear transformations of the dependent variable when there was a non-zero time effect: ``This problem occurs because nonlinear transformations of the dependent variable imply different marginal effects on the dependent variable at different levels of the dependent variable'' \cite[p.~155]{Meyer1995}. This warning was echoed more recently by \citet{KahnLang2020}.}

I then shift my attention to the singular aim of this paper: to explicate the consequences of the functional form assumption when the distribution of group outcomes in the pre-policy period differ substantively. These consequences can be stark. For a given aggregate time effect, I show that with a sufficiently large difference between group outcome distributions in the pre-policy period, it is possible for the level- and log-specifications of a DD design to yield DD parameter estimates of \textit{different signs}. Thus, one may conclude that a policy or intervention raised outcomes for the treatment group using one functional form, but may conclude the precise opposite using a different functional form, even with the same data, the same sample period, and the same control variables. To my knowledge, this is the first paper to methodically document this disparity. Given the wide use of DD designs for policy evaluation in areas that can give rise to such large differences in baseline outcome distributions -- e.g., gender gaps in earnings, race gaps in the length of incarceration spells, house prices across regions or states, or school test scores across different education regimes -- this point is likely to be of broad significance to applied researchers.

I provide a condition -- based on (i) the common time effect experienced by both groups, and (ii) the difference in relative baseline outcome means -- as to when we should expect a sign switch for the log-dependent variable case. This can straightforwardly be expressed in terms of the parameters of an additive DD model. Next, using first simulations, and then two empirical examples, I provide evidence of the sign disparity in estimated DD coefficients.\footnote{In order to maintain a singular focus on the consequences of functional form assumptions for DD designs, I restrict my attention to the non-staggered timing, binary treatment case for this paper.} I verify the condition I propose with simulation results.


In both empirical settings, using the same data and the same set of controls and fixed effects, I document a statistically significant DD estimate for a levels specification that is the opposite sign that that from a log specification. Both of these empirical case studies share the feature that the outcome distributions of treated and untreated groups are sufficiently different -- a necessary, but not sufficient condition to generate a sign difference in DD estimates for level and log specifications.

The first empirical example examines differences in total earnings for white and Black men before and after the Great Recession. In the aftermath of the Great Recession, average earnings for both groups fell. Due to considerably lower baseline earnings for Black men, while I estimate a positive DD estimate in levels, I estimate a statistically significant negative DD coefficient in logs. The reason for this is that the smaller fall in earnings for Black men was larger in proportional terms. Thus, a study investigating the differential racial impact of the Great Recession on male earnings would yield an opposing conclusion depending on whether the researcher chose a levels- or log-dependent variable specification.

In the second empirical case study, I consider differential price responses in the London housing market in the period around the Brexit vote. The hypothesis in this case is that the higher exposure of Inner London properties to overseas investors may have left this area more exposed to Brexit-related uncertainty than Outer London. This is another setting with large differences in the baseline outcome distribution across groups -- the average price of a house in Inner London is almost double that of Outer London. Consequently, while the levels of house prices rose significantly more in Inner London compared to Outer London, the growth rate of Outer London was larger, which again leads to a sign difference in DD parameter estimates in the level and log specifications.


This paper contributes to the difference-in-differences literature by making clear the consequences of functional form assumptions in DD designs, particularly when working with groups with large baseline differences in outcome distributions. This builds on work by both \cite{Meyer1995} and \citet{KahnLang2020} who noted that functional form assumptions would matter in such cases. This work also relates to the recent work by \citet{Roth2023}, who set out the conditions under which the parallel trend assumption is insensitive to functional form.

The remainder of the paper is organized as follows. Section \ref{sec:Model} provides an overview of the DD model, makes precise what the level- and log-specifications are estimating and provides a condition for when we will find a sign-switch between the two specifications. Section \ref{sec:Sim} provides simulation evidence to both verify the sign-switch condition, and to highlight cases that generate disparate coefficient estimates across specifications. Section~\ref{sec:EmpiricalCaseStudies} presents two empirical case studies where I document opposite-sign DD estimates for level- and log-outcome variable specifications. Section \ref{sec:Conclusion} concludes.

\section{\label{sec:Model}The DD Model}
Individual $i$ can belong to treatment group ($D_i=1$) or untreated comparison group ($D_i=0$).
We observe individuals in two periods -- $T_t=0$ and $T_t=1$. Those in treatment group receive treatment in period 1.
Using a potential outcomes framework, we can write down the realized outcome as $Y_{it} = (1-D_i)Y_{it}(0) + D_iY_{it}(1)$, where $Y_{it}(0)$ and $Y_{it}(1)$ are the potential outcomes for individual $i$  in absence of treatment and upon receipt of treatment respectively.
We write the DD estimand as:
\begin{align}
\alpha^{ATT} &= \left\{E[Y_{it} \mid D_i=1, T_t=1] - E[Y_{it} \mid D_i=1, T_t=0]\right] \nonumber \\
             &- \left[E[Y_{it} \mid D_i=0, T_t=1] - E[Y_{it} \mid D_i=0, T_t=0]\right\} 
\label{Eq:ATT} 
\end{align}
The sample analog of (\ref{Eq:ATT}) is the DD estimator
\begin{equation}
\hat{\alpha}^{DD} = \left[\overline{Y}_{T1} - \overline{Y}_{T0} \right] - \left[\overline{Y}_{C1} - \overline{Y}_{C0} \right]  \, ,
\label{Eq:DD1} 
\end{equation}
where the subscript T and C refer to treatment and control groups respectively. The subscripts 0 and 1 respectively refer to the pre- ($T_t=0$) and post-policy ($T_t=1$) periods.
A simple regression specification we can use to estimate the ATT parameter is:
\begin{equation}
Y_{it} = \alpha_1 + \alpha_2 Treat_i + \alpha_3 Post_t + \alpha_4 (Treat_i \times Post_t) + \epsilon_{it} \, ,
\label{Eq:DD2} 
\end{equation}
where $Treat_i$ is a treatment indicator, $Post_t$ the post-period indicator, and $\alpha_4$ is the parameter of interest.

Writing down an analogous specification with a log-transformed dependent variable of the form:
\begin{equation}
\ln Y_{it} = \beta_1 + \beta_2 Treat_i + \beta_3 Post_t + \beta_4 (Treat_i \times Post_t) + \mu_{it} \, ,
\label{Eq:logDD1} 
\end{equation}
implies an underlying model for $Y_{it}$ that is multiplicative rather than additive:
\begin{equation}
Y_{it} = \exp(\beta_1 + \beta_2 Treat_i + \beta_3 Post_t + \beta_4 (Treat_i \times Post_t)) \eta_{it} \, ,
\label{Eq:expDD1} 
\end{equation}
where $\mu_{it} = \ln \eta_it$. As noted by \citet{Mullahy1999}, and again by \citet{Ciani2019}, we can write an expression for the exponentiated DD parameter, based on the multiplicative model, as:
\begin{equation}
\exp(\beta_4) = \cfrac{\cfrac{E[Y_{it} \mid D_i=1, T_t=1]}{E[Y_{it} \mid D_i=1, T_t=0]}}{\cfrac{E[Y_{it} \mid D_i=0, T_t=1]}{E[Y_{it} \mid D_i=0, T_t=0]}} = \frac{g_T}{g_C} \, ,
\label{Eq:expDD2} 
\end{equation}
where $g_C$ and $g_T$ are the respective growth rates in the outcome for control and treated groups. 

Recalling that $\ln (1+z) \approx z$ for small values of $z$, and returning to Equation (\ref{Eq:logDD1}), we see that the DD parameter we estimate ($\beta_4$) with a log-dependent variable can be expressed as:
\begin{equation}
\beta_4 \approx \exp(\beta_4) -1 = 
\cfrac{
\cfrac{E[Y_{it} \mid D_i=1, T_t=1]}{E[Y_{it} \mid D_i=1, T_t=0]} - \cfrac{E[Y_{it} \mid D_i=0, T_t=1]}{E[Y_{it} \mid D_i=0, T_t=0]}
}{
\cfrac{E[Y_{it} \mid D_i=0, T_t=1]}{E[Y_{it} \mid D_i=0, T_t=0]}
} = \frac{g_T - g_C}{g_C} \, .
\label{Eq:logDD2} 
\end{equation}
Equation (\ref{Eq:logDD2}) makes clear that when we estimate a DD specification with a log-dependent variable (as in Equation (\ref{Eq:logDD1})), we are estimating an approximation of the  \textit{proportional difference in growth rates} of the outcome between the treated and untreated groups over the two periods. This is very different from what we measure with a level dependent variable -- the difference between groups in changes over time.

In Proposition~\ref{Prop:signswitch} below, I outline the conditions under which we will find a sign switch using a level and log specification.
\begin{proposition}
For a given (non-zero) aggregate time effect ($\alpha_3 \neq 0$), if outcomes means at baseline are sufficiently different in relative terms, the functional form decision of specifying a level- or log-dependent variable can yield a sign difference for the DD parameter. More specifically:\\

\noindent{}when $0 < \left|\Delta_T-\Delta_C\right|  < \left|\Delta_C \cfrac{(E[Y_{T0}]-E[Y_{C0}])}{E[Y_{C0}]}\right| \, ,$ we will have $sign(\alpha_4) \neq sign(\beta_4)$.\\ 

\noindent{}This condition may alternatively be expressed in terms of the parameters of the additive DD model as:\\
when $\left|\alpha_4\right|  < \left|\alpha_3 \cfrac{\alpha_2}{\alpha_1}\right| \, ,$ we will have $sign(\alpha_4) \neq sign(\beta_4)$. \\

\noindent{}Proof in Appendix~\ref{sec:proofs}.
\label{Prop:signswitch} 
\end{proposition}
The Proposition above uses the notation $E[Y_{C0}] = E[Y_{it} \mid D_i=0, T_t=0]$, $E[Y_{T0}] = E[Y_{it} \mid D_i=1, T_t=0]$, $\Delta_C = E[Y_{it} \mid D_i=0, T_t=1] - E[Y_{it} \mid D_i=0, T_t=0]$, and $\Delta_T = E[Y_{it} \mid D_i=1, T_t=1] - E[Y_{it} \mid D_i=1, T_t=0]$. In the simulation approach I detail in Section~\ref{sec:Sim}, I verify Proposition~\ref{Prop:signswitch}.

\paragraph{The Parallel Trends Assumption}
The key identifying assumption for the DD estimator to return the ATT is the parallel trends assumption. In recent work, \citet{Roth2023} set out the conditions under which the parallel trend is insensitive to functional form. These conditions -- either random assignment of treatment, stationarity of the distribution of potential outcomes for the untreated setting, $Y(0)$, or a combination of these two cases -- are considerably stricter than is typically assumed by empirical researchers using DD approaches. When such conditions are not met, one must choose, and consequently justify, a functional form for the outcome variable -- a parallel trend in levels obviates a parallel trend also holding in logs, and vice versa. This point was made by \citet{Meyer1995}, and has since been reiterated by several authors, including \citet{Angrist2009} and \cite{KahnLang2020}.

In one sense, these findings regarding functional form and the parallel trend assumption limit the scope of cases that may benefit from the insights of this paper. If (i) a parallel trend in levels precludes a parallel trend in logs and (ii) empirical researchers typically assess the validity of using DD methods by some form of pre-trend inspection and/or testing\footnote{See \citet{Roth2022} for a discussion on the pitfalls of such pre-testing}, then does the the singular aim of this paper -- to call attention to the possibility of a sign switch in DD estimates based on functional form assumptions --  carry any weight? I argue that it does, both in cases that satisfy the conditions set out by \citet{Roth2023}, or in cases where limited pre-policy data limits the statistical power to detect divergent pre-trends. Having a clearer sense of when such issues may arise will hopefully be of use to applied researchers.

\section{\label{sec:Sim}Simulation Results}
In this section, I provide simulation results to show that, using an additive and multiplicative model, one may estimate an ATT that differs not just in magnitude, but in sign. I take the additive model as the data generating process (DGP) in this case, and compare level- and log-dependent variable based DD specifications.\footnote{If one is interested in the complementary case where the data generating process is based on the multiplicative model, useful references include \citet{Silva2006} and \citet{Ciani2019}.}

The DGP is based on Equation (\ref{Eq:DD2}), with different simulation specifications using different parameters values for $\alpha_1$, $\alpha_2$, $\alpha_3$, and $\alpha_4$. In all cases, $E[\epsilon_{it} \mid D_i, T_t] = 0$ and $\sigma_{\epsilon} = .2$. The proportion of treated and the proportion in the post period are .5 in both cases, and the total sample size is 40,000. This means each of the $2\times2$ DD cells has 10,000 observations.

The simulation results below reflect the key insight of this paper --because the DD parameter from a level- and log-dependent variable specification respectively reflect a difference in differences in levels and a proportional difference in growth rates, holding fixed the time effect, one just needs to shift the distribution of outcomes for one of the groups to drive a wedge between the resulting parameter estimates.

\begin{center}
  \begin{table}[htb] \centering
\newcolumntype{C}{>{\centering\arraybackslash}X}

\caption{\label{tab:simulations_table_1}Simulating a Postive Levels DD Effect That Yields a Zero or  Negative Log Effect}
{\footnotesize
\begin{tabularx}{\linewidth}{lCCCCC}

\toprule
\multicolumn{1}{c}{ }& \multicolumn{3}{c}{{\textbf{Zero Log Effect}}} &  \multicolumn{2}{c}{\textbf{Negative Log Effect}} \tabularnewline  \cmidrule(l{2pt}r{5pt}){2-4} \cmidrule(l{2pt}r{5pt}){5-6}  \addlinespace[-2ex] \tabularnewline
{}&{(1)}&{(2)}&{(3)}&{(4)}&{(5)} \tabularnewline
\midrule \addlinespace[\belowrulesep]
\textbf{Level-Dependent Variable:}&&&&& \tabularnewline
DD Estimate [\(\hat{\alpha}_4\)]&0.40***&1.00***&0.40***&1.00***&0.80*** \tabularnewline
&(0.00)&(0.00)&(0.00)&(0.00)&(0.00) \tabularnewline
\textbf{Log-Dependent Variable:}&&&&& \tabularnewline
DD Estimate [\(\hat{\beta}_4\)]&--0.000&--0.000&--0.000&--0.043***&--0.017*** \tabularnewline
&(0.000)&(0.000)&(0.000)&(0.000)&(0.001) \tabularnewline
exp(\(\hat{\beta}_4\))--1&--0.000&--0.000&--0.000&--0.042***&--0.017*** \tabularnewline
&(0.000)&(0.000)&(0.000)&(0.000)&(0.001) \tabularnewline
\addlinespace[.5ex] \midrule \addlinespace[1ex] \(\overline{Y}_{C0}\)&10.00&10.00&20.00&10.00&5.00 \tabularnewline
\(\overline{Y}_{C1}\)&12.00&15.00&22.00&12.00&6.00 \tabularnewline
\addlinespace[.25ex] \(\overline{Y}_{T0}\)&12.00&12.00&24.00&20.00&10.00 \tabularnewline
\(\overline{Y}_{T1}\)&14.40&18.00&26.40&23.00&11.80 \tabularnewline
\((g_T - g_C)/g_C\)&--0.000&0.000&--0.000&--0.042&--0.017 \tabularnewline
\bottomrule \addlinespace[\belowrulesep]

\end{tabularx}
\begin{flushleft}
\scriptsize \textbf{Notes}: Results based on 10000 simulation runs. DD estimates for both a level- and log-dependent variable are presented in each column. The tables display the mean and (in parentheses) bootrapped standard error of the DD estimates across all simulation runs. At the base of the table, the four elements of the DD are presented for reference. The sample size is 40000 in each simulation. Proportion treated and proportion in the post period are .5 and .5 respectively. The standard deviation of the error term is .2 in all simulations.
\end{flushleft}
}
\end{table}

\end{center}

Table~\ref{tab:simulations_table_1} presents the first set of simulation results. All DD estimates for the level-dependent variables are positive, whereas the estimates for the log-dependent variable specifications are either precisely zero, or negative. 
For log specifications, I present both $\beta_4$, and $\exp(\beta_4) -1$, of which the latter equals the proportional difference in growth rates (see Equation~(\ref{Eq:logDD2})).
At the base of the table, I present the cell means for the outcome variable across the four DD groups -- thus one can easily see how the parameters are generated -- as well as the proportional growth rates that I calculate from these cell means.\footnote{Given that I do not introduce additional control variables in the simulations, the underlying model is saturated -- the four parameters of Equation (\ref{Eq:DD2}) map directly to the four cells of the DD design. This means that the proportional growth rate I present in the final row of the table matches perfectly $\exp(\beta_4)-1$.}

The key point of this table is to show that one can generate a positive DD estimate from a levels specification and, by merely shifting the baseline outcome distribution of the untreated group, also generate a precise zero or a negative DD estimate from a log specification. In Table~\ref{tab:simulations_table_2} I present a complementary set of simulation results, where I fix the level specification to yield a DD estimate that is precisely zero (once again, the cells sample averages at the base of the table provide the key insight into how this is operationalized) for a levels specification, but which yields either a negative or positive DD estimate from a log specification.

\begin{figure}[htb]
  \centering
    \caption{The Sign Switch Occurs When $\alpha_4 = \alpha_3 \cfrac{\alpha_2}{\alpha_1}$}
    \begin{subfigure}[b]{0.48\linewidth}
      \includegraphics[width=\linewidth]{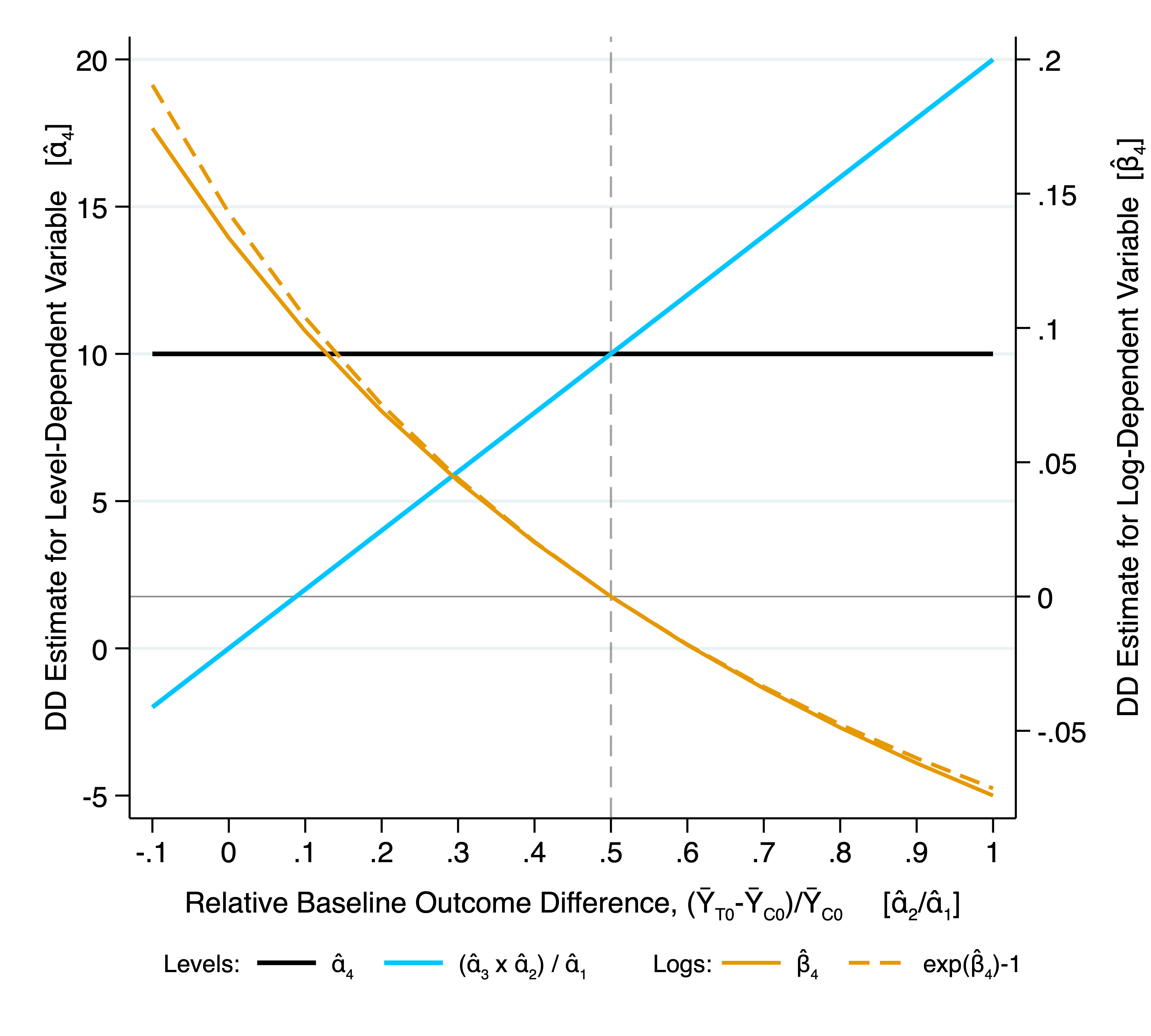}
      \caption{Varying Baseline Outcome Differences}
      \label{fig:simulations_graphs_baseline}
    \end{subfigure}
    \hspace{-10pt}
    \begin{subfigure}[b]{0.48\linewidth}
      \includegraphics[width=\linewidth]{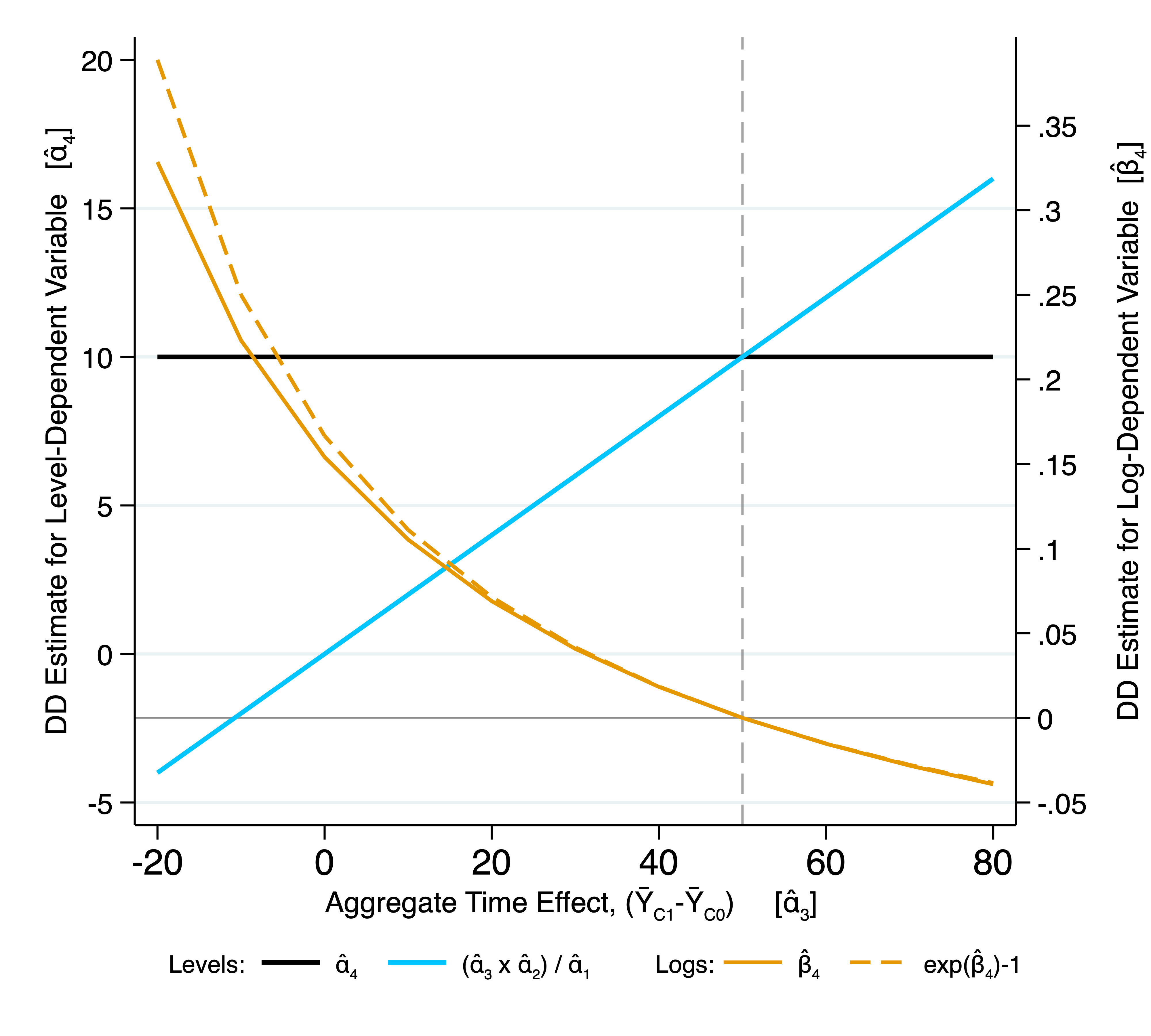}
      \caption{Varying the Aggregate Time Effect}
      \label{fig:simulations_graphs_timeFE}
    \end{subfigure}
  \captionsetup{belowskip=-3pt}
  \caption*{{\scriptsize \textbf{Notes:} The underlying DGP that creates both graphs is an additive DD model specification, thus taking the form of Equation~(\ref{Eq:DD2}). In the left-hand graph, I set the following as fixed: ${Y}_{C0} = 50$, ${Y}_{T1} - {Y}_{T0} = 30$, ${Y}_{C1} - {Y}_{C0} = 20$. What varies is the relative baseline difference: $({Y}_{T0} - {Y}_{C0})/{Y}_{C0} = \alpha_2/\alpha_1$. Once the relative baseline difference is fixed, this yields a specific value of ${Y}_{T0}$ for the simulation exercise. In the right-hand graph, I set the following as fixed: ${Y}_{C0} = 50$, ${Y}_{T0} = 60$, $({Y}_{T1} - {Y}_{T0}) = ({Y}_{C1} - {Y}_{C0}) +10$. What varies is the aggregate time effect, $({Y}_{C1} - {Y}_{C0})$. Once the aggregate time effect is fixed, this yields a specific value for both ${Y}_{C1}$ and ${Y}_{T1}$ for the simulation exercise. For each increment of relative baseline difference and aggregate time effect values respectively, 10,000 simulations are conducted. In both cases, the specification leads to a constant (levels) DD effect of 10 units, which is why the black line is horizontal. The sample size is 40,000 in each simulation. Proportion treated and proportion in the post period are .5 and .5 respectively. The standard deviation of the error term is .2 in all simulations.}}
  \label{fig:simulations_graphs}
\end{figure}
In Figure~\ref{fig:simulations_graphs}, I present DD estimates from both a levels-\footnote{These are presented using the left-hand $y$-axis, black line.}  and log-based\footnote{These are presented using the right-hand $y$-axis, orange lines.} specification for two separate simulation experiments. Based on the parameters I choose, the DD coefficient estimate is a constant equal to 10 for the levels specification. Keeping all parameters fixed except for either the relative baseline outcome difference (Figure~\ref{fig:simulations_graphs_baseline}) or the aggregate time effect (Figure~\ref{fig:simulations_graphs_timeFE}), I verify Proposition~\ref{Prop:signswitch}. The figures make clear two key points. First, keeping the levels-based DD estimate fixed at a constant value and altering either the relative difference in baseline means ($\alpha_2/\alpha_1$) or the aggregate time effect ($\alpha_3$), it is possible to generate either a negative or a positive DD estimate for the log-dependent variable specification. Second, the sign switch occurs at precisely the point stated in Proposition~\ref{Prop:signswitch}.

\begin{figure}[htb]
  \centering
    \caption{There is no Sign Switch Without a Baseline Outcome Difference or a non-Zero Aggregate Time Effect}
    \begin{subfigure}[b]{0.48\linewidth}
      \includegraphics[width=\linewidth]{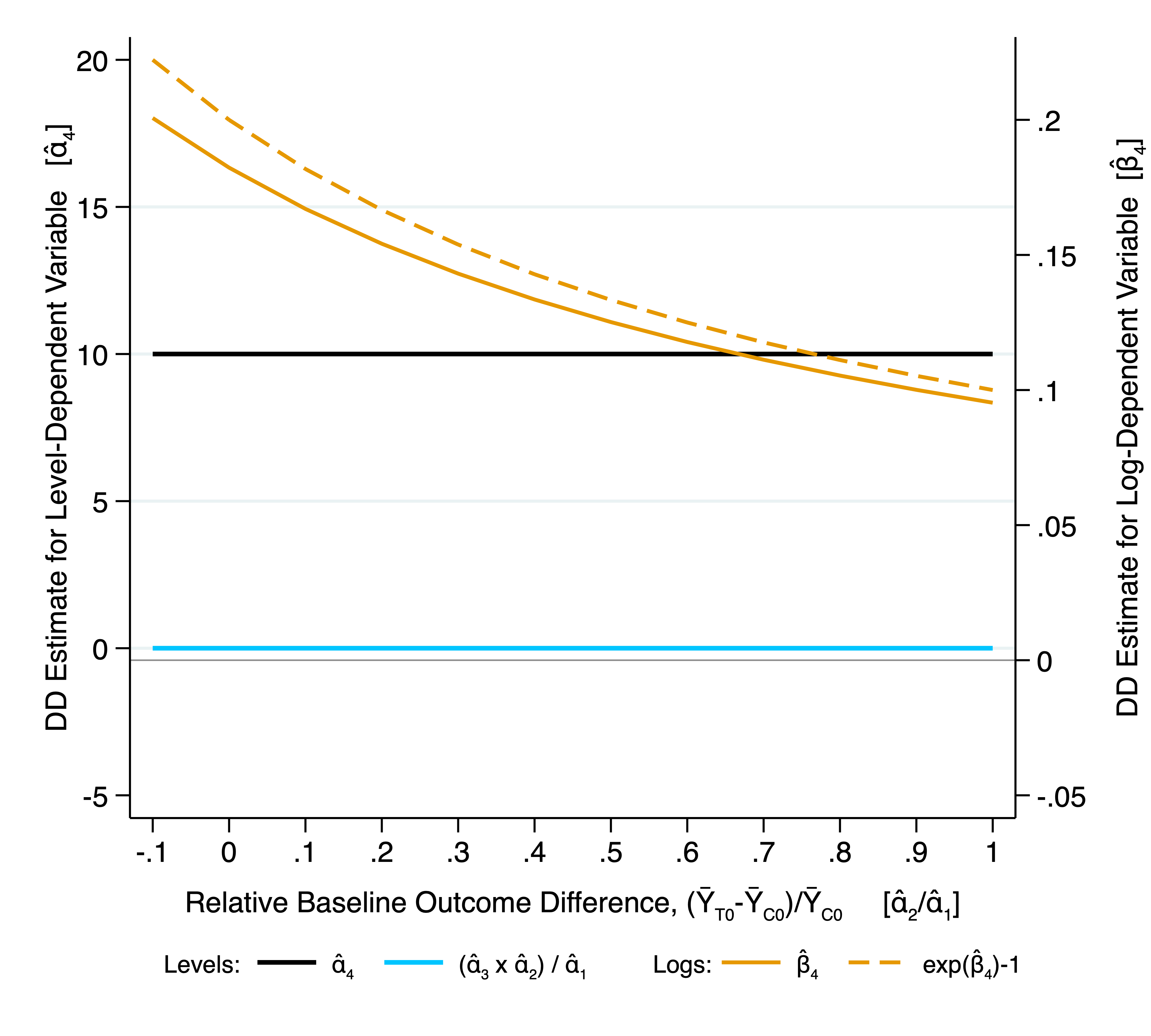}
      \caption{Varying Baseline Outcome Differences}
      \label{fig:simulations_graphs_baseline_zero}
    \end{subfigure}
    \hspace{-10pt}
    \begin{subfigure}[b]{0.48\linewidth}
      \includegraphics[width=\linewidth]{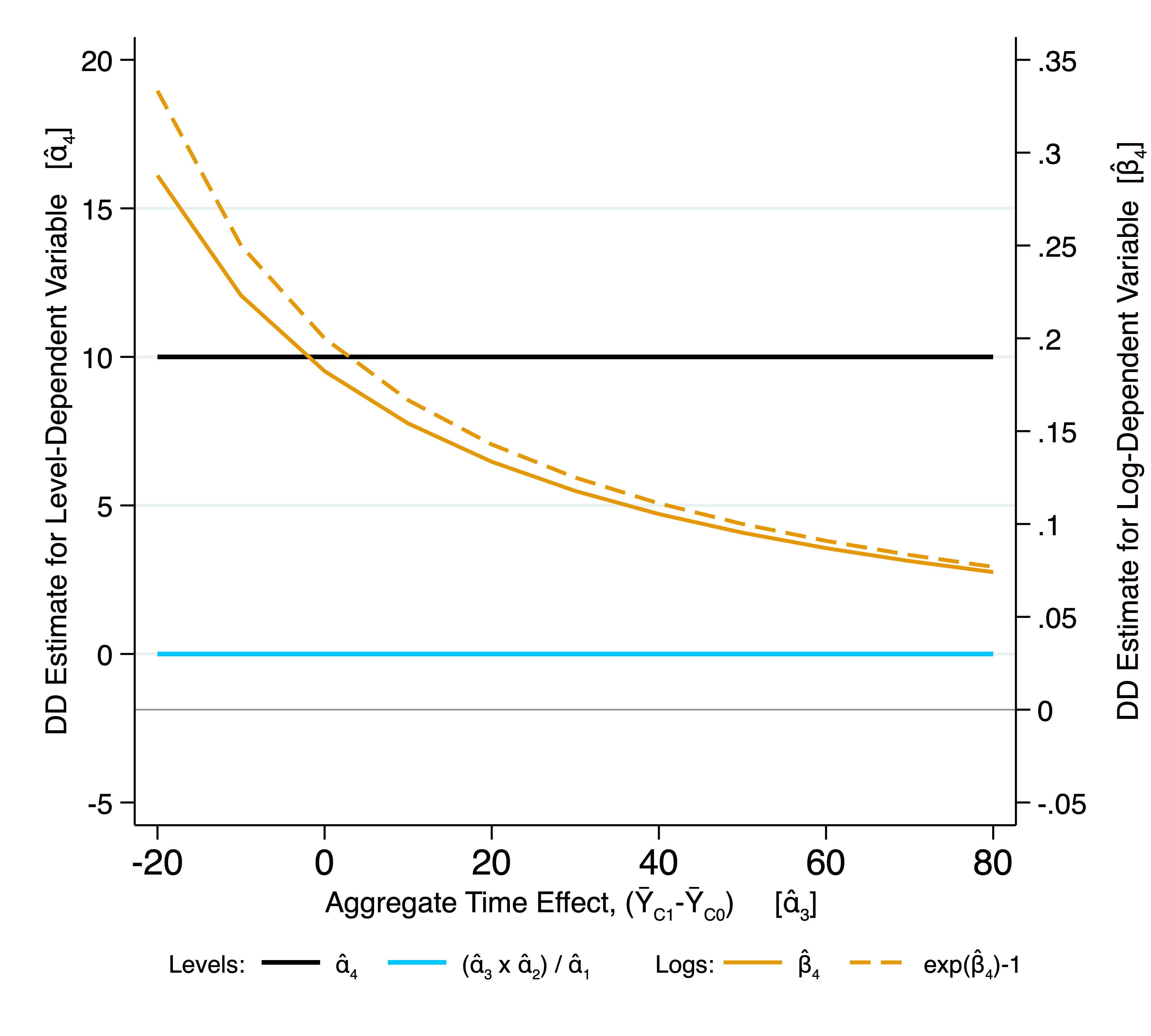}
      \caption{Varying the Aggregate Time Effect}
      \label{fig:simulations_graphs_timeFE_zero}
    \end{subfigure}
  \captionsetup{belowskip=-3pt}
  \caption*{{\scriptsize \textbf{Notes:} The underlying DGP that creates both graphs is an additive DD model specification, thus taking the form of Equation~(\ref{Eq:DD2}). In the left-hand graph, I set the following as fixed: ${Y}_{C0} = 50$, ${Y}_{T1} - {Y}_{T0} = 30$, ${Y}_{C1} - {Y}_{C0} = 0$ (this last choice means the aggregate time effect is zero). What varies is the relative baseline difference: $({Y}_{T0} - {Y}_{C0})/{Y}_{C0} = \alpha_2/\alpha_1$. Once the relative baseline difference is fixed, this yields a specific value of ${Y}_{T0}$ for the simulation exercise. In the right-hand graph, I set the following as fixed: ${Y}_{C0} = 50$, ${Y}_{T0} = 50$, $({Y}_{T1} - {Y}_{T0}) = ({Y}_{C1} - {Y}_{C0}) +10$. Note, there is no longer a baseline difference in outcome means. What varies is the aggregate time effect, $({Y}_{C1} - {Y}_{C0})$. Once the aggregate time effect is fixed, this yields a specific value for both ${Y}_{C1}$ and ${Y}_{T1}$ for the simulation exercise. For each increment of relative baseline difference and aggregate time effect values respectively, 10,000 simulations are conducted. In both cases, the specification leads to a constant (levels) DD effect of 10 units, which is why the black line is horizontal. The sample size is 40,000 in each simulation. Proportion treated and proportion in the post period are .5 and .5 respectively. The standard deviation of the error term is .2 in all simulations.}}
  \label{fig:simulations_graphs_zero}
\end{figure}
The condition provided in Proposition~\ref{Prop:signswitch} indicates that we require both a non-zero aggregate time effect ($\alpha_3 \neq 0$) and a difference in baseline outcome means ($\alpha_2 \neq 0$) to find a sign switch across level- and log-specifications. The purpose of Figure~\ref{fig:simulations_graphs_zero} is to show that this is indeed the case. The underlying DGPs are identical to those underlying Figure~\ref{fig:simulations_graphs}, except for the key difference that in Figure~\ref{fig:simulations_graphs_baseline_zero} I set the aggregate time effect to zero, and in Figure~\ref{fig:simulations_graphs_timeFE_zero} I set the baseline outcome mean wedge to zero. As one can see, the log-based DD estimate (shown in orange) never crosses the zero line i.e., there is no sign switch.

Before moving to the empirical case studies, it is worth reflecting on what we have learned so far. What is concerning for applied researchers is that, if one has identified (i) a particular treatment and control group in order to evaluate a policy and (ii) a given sample period, then both the baseline outcome distributions of the respective groups and the resulting time effect is given. Thus, with a given sample, one may find oneself at the equivalent of either the left-hand side or the right-hand side of the $x$-axis in the graphs shown in  Figure~\ref{fig:simulations_graphs}. Unlike the simulation exercise presented here, this will not be a choice, and one will not be able to shift the ratio of relative baseline outcome means or the aggregate time effect to explore the sensitivity of the parameter estimates to functional form assumptions. I build on this point with the two empirical case studies below.

\section{\label{sec:EmpiricalCaseStudies}Empirical Case Studies}

I now present two empirical case studies. Both of the settings were chosen with an eye on the disparate baseline outcome distribution across the groups of interest, which as I note above is a necessary, but not sufficient, condition to generate DD estimates of opposing signs for the level- and log-dependent variable specifications. I first examine the differential racial impact of the great recession on working age males in the US. I then investigate the differential impact of the Brexit vote on Inner vs Outer London property prices.

\subsection{\label{sec:EmpiricalCaseStudiesGR}Male Earnings and The Great Recession}

The empirical specification I use to investigate the differential racial impact of the Great Recession on male earnings is:
\begin{equation}
  Income_{it} = \delta Black_i +  \gamma (Black_i \times Post_t) + X_{i}^{'} \pi + \theta_{s\times t} + \epsilon_{it}
  \label{Eq:GRDD1}
\end{equation}
%
where $Income_{it}$ is the total income earned in the previous year, and is specified in either levels or logs, for individual $i$ in year $t$. $Black_i$ takes a value 0 for non-Hispanic white males, and a value of 1 for Black males. $Post_t$ is a dummy for the post-Great Recession period\footnote{Given that income in year $t$ reflects income from the previous year, I code $Post_t = \mathbbm{1}[Year>=2009]$ in order to capture the Great Recession kicking in in 2008.}. $X_i$ is a vector of individual characteristics that includes dummies for highest level of educational attainment, dummies for potential experience in years, an indicator for being married, and dummies for metro area classification. $\theta_{s\times t}$ is a set of state-by-year fixed effects. The error term is $\epsilon_{it}$. I specify Eicker-Huber-White standard errors throughout. I present a set of summary statistics for the setting, and an overview of the data and sample selection decisions, in Appendix~\ref{sec:summstatsGR}.

\begin{center}
  \begin{table}[htb] \centering
\newcolumntype{C}{>{\centering\arraybackslash}X}

\caption{\label{tab:an_DD_CPS_1}Level- and Log-Specifications Yield Estimates of Opposing Signs for the Impact of the Great Recession on the Racial Income Gap}
{\footnotesize
\begin{tabularx}{\linewidth}{lCCCCCC}

\toprule
&{(1)}&{(2)}&{(3)}&{(4)}&{(5)}&{(6)} \tabularnewline \midrule
\multicolumn{1}{c}{ } & \multicolumn{2}{c}{{\textbf{All Education Levels}}} & \multicolumn{2}{c}{{\textbf{No College}}} &  \multicolumn{2}{c}{\textbf{College}} \tabularnewline  \cmidrule(l{2pt}r{5pt}){2-3} \cmidrule(l{2pt}r{5pt}){4-5} \cmidrule(l{2pt}r{5pt}){6-7}  \addlinespace[-2ex] \tabularnewline
{}&{Income in Levels}&{Income in Logs}&{Income in Levels}&{Income in Logs}&{Income in Levels}&{Income in Logs} \tabularnewline
\midrule \addlinespace[\belowrulesep]
Post \(\times\) Black &775*&-.0361**&1045&-.0231&346&-.0511** \tabularnewline
&(467)&(.0142)&(762)&(.0191)&(487)&(.0213) \tabularnewline
\addlinespace[.2ex] exp(\(\hat{\beta}_4\))--1&&-.0354***&&-.0228&&-.0498** \tabularnewline
&&(.0137)&&(.0187)&&(.0202) \tabularnewline
\addlinespace[2ex] \midrule \addlinespace[1ex] \(\overline{Y}_{C0}\)&48088&48088&58498&58498&31485&31485 \tabularnewline
\addlinespace[.1ex] \(\overline{Y}_{C1}\)&45905&45905&55160&55160&29806&29806 \tabularnewline
\addlinespace[.1ex] \(\overline{Y}_{T0}\)&30020&30020&38644&38644&22197&22197 \tabularnewline
\addlinespace[.1ex] \(\overline{Y}_{T1}\)&28495&28495&36080&36080&20866&20866 \tabularnewline
\addlinespace[.5ex] Adjusted \(R^2\)&.201&.202&.159&.152&.0846&.123 \tabularnewline
Observations&253,491&253,491&155,947&155,947&97,544&97,544 \tabularnewline
\bottomrule \addlinespace[\belowrulesep]

\end{tabularx}
\begin{flushleft}
\scriptsize \textbf{Notes}: *** denotes significance at 1\%, ** at 5\%, and * at 10\%. Eicker-Huber-White standard errors in parentheses. The dependent variable is total income  in 1999 Dollars, measured in levels in odd-numbered columns, and in logs in even-numbered columns. The following control variables are included in all regressions: state-by-year fixed effects, dummies for highest level of educational attainment, dummies for potential experience in years, an indicator for being married, and dummies for metro area classification. Data used: CPS 2005-2012.
\end{flushleft}
}
\end{table}

\end{center}
%
Table~\ref{tab:an_DD_CPS_1} presents the key parameter estimates from both a level and log version of Equation~(\ref{Eq:GRDD1}). The DD estimate presented in Column (1) suggests that Black men fared slightly better in the Great Recession than their white counterparts. Figure~\ref{fig:GRtrends} shows that both groups suffered absolute falls in incomes during this period, so the DD estimate in Column (1) reflects that the income drop for Black men was smaller than that for white men. The results in Column (2) present a starkly different conclusion, with a statistically significant, negative DD estimate. Using the raw means for the four DD cells, we can reconcile these two estimates -- although Black men experience a lower absolute income drop, relatively it was larger than for white men as the drop occurred from a lower baseline level of income, which is why the coefficient in Column (1) is positive, and in Column (2) is negative.

With the benefit of having worked through what one recovers from a level and log specification in Section~\ref{sec:Model}, and seen the consequences for these different functional form specifications of disparate baseline outcome distributions in Section~\ref{sec:Sim}, with the information at hand it is a straightforward task to dissect the reason for the sign difference documented in Columns (1) and (2). However, in a typical applied setting, where a researcher is aiming to document the impact of a new policy using a DD model, the source of such disparate results may be less clear. It is the hope that this paper will aide in such settings.

The remaining columns of Table~\ref{tab:an_DD_CPS_1} present the DD parameters for both level and log specification, splitting the sample by education levels.

\subsection{\label{sec:EmpiricalCaseStudiesBrexit}London House Prices and the Brexit Vote}

I next turn to a different setting -- the London housing market in the period around the Brexit vote. To understand the differential impact on house prices in Inner and Outer London, I specify a hedonic house price model of the form:
\begin{align}
  Price_{it} =& \delta (Inner_i \times Post_t) + \sum_{m=1}^M (Market_m \times Post_t \times X_{i}^{'} \gamma_m)  + \pi_{m \times t} + \theta_{b} + \epsilon_{it} \text{ ,}
  \label{Eq:HPDD1}
\end{align}
where $Price_{it}$ is the house price of house $i$ (specified in either levels or logs), sold in period $t$ (measured at the month-by-year level). $Inner$ takes the value of 0 for Outer London boroughs, and the value of 1 for Inner London boroughs.\footnote{To confuse matters there are two definitions of Inner London -- the statutory definition, and the statistical version. In order to be consistent with both measures, I apply the strictest definition -- in order for me to classify a borough as Inner London, a borough must be classified as Inner London by both definitions. This leads me to code the following boroughs as belonging to Inner London: Camden, Hackney, Hammersmith and Fulham, Islington, Kensington and Chelsea, Lambeth, Lewisham, Southwark, Tower Hamlets, Wandsworth, and Westminster.} $Post_t$ is a dummy for properties sold post-Brexit vote, and $\delta$ is the parameter of interest. 

$X_{i}$ is a vector of property characteristics, specifically interactions between dummies for property type categories and new build status, and interactions between dummies for leasehold and new build status. I include interaction between the vector of housing characteristics, $X_i$, and market dummies in order to respect the ``law of one price function'' \citep{Bishop2020}. This allows the valuation of key property characteristics to vary across local housing markets. I allow the coefficients on all housing characteristics to differ in the pre and post periods, thereby allowing the hedonic price function to shift post-policy. I do so in order to avoid conflation bias \citep{Kuminoff2014,Banzhaf2021}. 

$\pi_{m \times t}$ captures month-by-year housing market shocks to house prices. Housing markets are Travel To Work Areas -- similar to Commuting Zones in the US. $\theta_b$ is a spatial fixed effect at the level of Output Area, akin to a census block in the US. Output Areas (OA) are the smallest census-based geographical unit -- there are 181,408 of these in England and Wales, with an average population of 309 at the 2011 census.\footnote{\url{https://www.ons.gov.uk/peoplepopulationandcommunity/populationandmigration/populationestimates/bulletins/2011censuspopulationandhouseholdestimatesforsmallareasinenglandandwales/2012-11-23}}. The Output Area fixed effect will capture all time-invariant local amenities -- green spaces, transport links, shops, proximity to busy roads or motorways, as well as many slow-moving time-varying area characteristics (I am considering a minimum of 2 years, and a maximum of 4 years for these estimations), such as access to good schools or proximity to sources of pollution. 
The error term is $\epsilon_{it}$. I specify Eicker-Huber-White standard errors throughout.
\begin{center}
  \begin{table}[htb] \centering
\newcolumntype{C}{>{\centering\arraybackslash}X}

\caption{\label{tab:an_DD_HP_pt04_price_1}Level- and Log-Specifications Yield Estimates of Opposing Signs for the Impact of the Brexit Vote on Inner London House Prices}
{\footnotesize
\begin{tabularx}{\linewidth}{lCCCCC}

\toprule
\multicolumn{1}{c}{ }& \multicolumn{5}{c}{\textbf{Time Window Around Brexit Vote}} \tabularnewline  \cmidrule(l{2pt}r{5pt}){2-6}  \addlinespace[-2ex] \tabularnewline
{}&{12 Months}&{15 Months}&{18 Months}&{21 Months}&{24 Months} \tabularnewline
\midrule \addlinespace[\belowrulesep]
\textbf{A.) All Property Types}&&&&& \tabularnewline
\midrule \textbf{Ai.) Price Specification}&&&&& \tabularnewline
\addlinespace[1ex] Post \(\times\) Inner London [\(\hat{\alpha_4}\)]&10,045&16,861**&12,125*&10,639*&4,588 \tabularnewline
&(8,285)&(7,031)&(6,416)&(6,035)&(5,713) \tabularnewline
\addlinespace[2ex] \textbf{Aii.) log(Price) Specification}&&&&& \tabularnewline
\addlinespace[1ex] Post \(\times\) Inner London [\(\hat{\beta_4}\)]&-.039***&-.0421***&-.0502***&-.0574***&-.0694*** \tabularnewline
&(.00379)&(.00334)&(.00307)&(.00286)&(.00267) \tabularnewline
\addlinespace[.2ex] exp(\(\hat{\beta_4}\))--1&-.0383***&-.0412***&-.049***&-.0558***&-.0671*** \tabularnewline
&(.00365)&(.0032)&(.00292)&(.0027)&(.00249) \tabularnewline
\addlinespace[2ex]  \(\overline{Y}_{C0}\)&457,259&451,339&445,855&439,377&437,491 \tabularnewline
\addlinespace[.1ex] \(\overline{Y}_{C1}\)&482,338&488,284&490,084&490,799&491,442 \tabularnewline
\addlinespace[.1ex] \(\overline{Y}_{T0}\)&791,380&779,916&776,787&773,525&775,656 \tabularnewline
\addlinespace[.1ex] \(\overline{Y}_{T1}\)&825,855&829,664&836,659&845,105&847,116 \tabularnewline
Observations&210,626&262,937&309,348&358,806&411,928 \tabularnewline
\addlinespace[2ex] \midrule  \textbf{B.) Apartments Only}&&&&& \tabularnewline
\midrule \textbf{Bi.) Price Specification}&&&&& \tabularnewline
\addlinespace[1ex] Post \(\times\) Inner London [\(\hat{\alpha_4}\)]&14,069**&12,321**&14,122***&11,360**&8,373* \tabularnewline
&(6,310)&(5,640)&(5,205)&(4,966)&(4,741) \tabularnewline
\addlinespace[2ex] \textbf{Bii.) log(Price) Specification}&&&&& \tabularnewline
\addlinespace[1ex] Post \(\times\) Inner London [\(\hat{\beta_4}\)]&-.0355***&-.0416***&-.0483***&-.0565***&-.0665*** \tabularnewline
&(.00473)&(.00418)&(.00383)&(.00355)&(.00331) \tabularnewline
\addlinespace[.2ex] exp(\(\hat{\beta_4}\))--1&-.0349***&-.0408***&-.0472***&-.055***&-.0643*** \tabularnewline
&(.00457)&(.00401)&(.00365)&(.00336)&(.0031) \tabularnewline
\addlinespace[2ex]  \(\overline{Y}_{C0}\)&340,331&336,618&333,686&328,396&325,193 \tabularnewline
\addlinespace[.1ex] \(\overline{Y}_{C1}\)&371,619&373,860&376,044&376,540&377,647 \tabularnewline
\addlinespace[.1ex] \(\overline{Y}_{T0}\)&674,757&665,064&659,124&656,169&652,712 \tabularnewline
\addlinespace[.1ex] \(\overline{Y}_{T1}\)&714,726&712,594&724,406&734,085&734,700 \tabularnewline
Observations&118,272&147,786&173,710&201,350&231,266 \tabularnewline
\bottomrule \addlinespace[\belowrulesep]

\end{tabularx}
\begin{flushleft}
\scriptsize \textbf{Notes}: *** denotes significance at 1\%, ** at 5\%, and * at 10\%. Eicker-Huber-White standard errors in parentheses. The dependent variable house price in specification i.) and log house price in specification ii.). The following control variables are included in all regressions: block fixed effects, post-by-TTWA-by-property-type-by-new-build dummies,  post-by-TTWA-by-leasehold-by-new-build dummies, and TTWA-by-year-by-month fixed effects. Data used: Land Registry Price Paid data 2014-2018.
\end{flushleft}
}
\end{table}

\end{center}
%
I first provide results for all property types in panel A of Table~\ref{tab:an_DD_HP_pt04_price_1}. The DD estimates for the level specification goes against my initial hypothesis that post-Brexit, Inner London property prices would suffer. For a variety of time windows around the Brexit vote ranging from 12 to 24 months, I document positive increases in Inner London prices relative to Outer London. As before, the log specification results are of the opposite sign, documenting a 4-7\% decline in relative growth rates of Inner London properties. As in the previous section, we are sufficiently informed with the cell sample means, and the trends documented in Figure~\ref{fig:BrexittrendsAll}, to understand why. Once again, the disparate baseline outcome distributions play a key role. While Inner London properties experience a slight increase in levels compared to Outer London properties, Inner London properties (which started off at a much higher baseline level) grew less, leading to a negative proportional difference in growth rates -- what the DD parameter approximates in a log specification.

Given that apartments account for almost 80\% of property transactions in Inner London (see Table~\ref{tab:balance_table_pricepaid_1}), in panel B of Table~\ref{tab:an_DD_HP_pt04_price_1} I restrict the sample to only apartment transactions, and repeat the analysis. The reason to do so is to get a fairer sense of the house price impact of Brexit. The unintended consequence of this sample restriction was to create a larger wedge between the baseline outcome distributions. Looking at Column (3), the ratio of sample means for treatment to control in the baseline period is 1.74 in panel A, but 1.98 in panel B. This increased disparity in baseline outcome sample means explains at least part of why the difference between the level and log specifications is even more pronounced in panel B.

\section{\label{sec:Conclusion}Conclusion}

The aim of this paper is to make clear the consequence of functional form assumptions when one uses a DD model in an empirical setting where the baseline outcome distribution across groups differs substantively. I provide a condition, based on the aggregate time effect and the relative difference in baseline means, whereby a level- and log-specification will yield estimates of the DD term of opposing signs. The key reason that this sign-switch can occur is that using a DD model with a log-dependent variable leads to the estimation not of a difference-in-differences, but rather an approximation of the relative difference in growth rates across groups.

Using both simulations and empirical examples, I show that one can obtain DD estimates of different signs depending on whether one specifies the outcome variable in levels or in logs. Given the wide use of DD models for policy evaluation in areas that can give rise to such large differences in baseline outcome distributions -- e.g., gender gaps in earnings, race gaps in the length of incarceration spells, house prices across regions or states, or school test scores across different education regimes -- this point is likely to be of broad significance to applied researchers.
\clearpage
\bibliographystyle{ecta}
\bibliography{logDD.bib,QJE_DD_2018_2022_v3.bib}

\newpage

\appendix
\section*{\centering{\huge{Appendix}}}
\bigskip

\beginappendixA
\section{\label{sec:proofs}Proofs}
Repeating elements of Section~\ref{sec:Model} to keep this section self-contained, we can write the DD estimand as:
\begin{align}
\alpha^{ATT} &= \left\{E[Y_{it} \mid D_i=1, T_t=1] - E[Y_{it} \mid D_i=1, T_t=0]\right] \nonumber \\
             &- \left[E[Y_{it} \mid D_i=0, T_t=1] - E[Y_{it} \mid D_i=0, T_t=0]\right\} 
\label{Eq:appATT} 
\end{align}
The sample analog of (\ref{Eq:ATT}) is the DD estimator
\begin{equation}
\hat{\alpha}^{DD} = \left[\overline{Y}_{T1} - \overline{Y}_{T0} \right] - \left[\overline{Y}_{C1} - \overline{Y}_{C0} \right] 
\label{Eq:appDD1} 
\end{equation}
A simple regression specification we can use to estimate the ATT parameter is:
\begin{equation}
Y_{it} = \alpha_1 + \alpha_2 Treat_i + \alpha_3 Post_t + \alpha_4 (Treat_i \times Post_t) + \epsilon_{it} \, ,
\label{Eq:appDD2} 
\end{equation}
where $Treat_i$ is a treatment indicator, $Post_t$ the post-period indicator, and $\alpha_4$ is the parameter of interest.

In Section~\ref{sec:Model}, I note that the DD parameter we estimate ($\beta_4$) with a log-dependent variable can be expressed as:
\begin{equation}
\beta_4 \approx \exp(\beta_4) -1 = 
\cfrac{
\cfrac{E[Y_{it} \mid D_i=1, T_t=1]}{E[Y_{it} \mid D_i=1, T_t=0]} - \cfrac{E[Y_{it} \mid D_i=0, T_t=1]}{E[Y_{it} \mid D_i=0, T_t=0]}
}{
\cfrac{E[Y_{it} \mid D_i=0, T_t=1]}{E[Y_{it} \mid D_i=0, T_t=0]}
} = \frac{g_T - g_C}{g_C} \, ,
\label{Eq:applogDD2} 
\end{equation}
where $g_C$ and $g_T$ are the respective growth rates in the outcome for control and treated groups.

\subsection{\label{sec:proofsProp1}Proof for Proposition~\ref{Prop:signswitch}}
First, when the level-DD is positive i.e.,  $\alpha_4>0$:
\begin{align}
\alpha_4 > 0 &\Rightarrow (E[Y_{11}] - E[Y_{10}]) - (E[Y_{01}] - E[Y_{00}]) >0   \nonumber \\
             &\Rightarrow  \Delta_1  -  \Delta_0  >0             \label{Eq:applevelwork1} 
\end{align}

\noindent{}Now consider the case where log-DD is zero i.e.,  $\beta_4=0$. This occurs when $g_T = g_C$ (see Equation~(\ref{Eq:applogDD2}))
\begin{align}
 \beta_4 = \exp(\beta_4) -1 =0 &\Rightarrow \left. \cfrac{E[Y_{11}]}{E[Y_{10}]} \middle/ \cfrac{E[Y_{01}]}{E[Y_{00}]}\right. - 1 =0 \nonumber \\
             &\Rightarrow \left. \cfrac{E[Y_{11}]}{E[Y_{10}]} \middle/ \cfrac{E[Y_{01}]}{E[Y_{00}]}\right. = 1 \nonumber \\
            &\Rightarrow \cfrac{E[Y_{11}]}{E[Y_{10}]}  = \cfrac{E[Y_{01}]}{E[Y_{00}]} \nonumber \\
\text{(minus 1 from both sides)} \qquad             &\Rightarrow \cfrac{E[Y_{11}]}{E[Y_{10}]} -1 = \cfrac{E[Y_{01}]}{E[Y_{00}]} -1 \nonumber \\
&\Rightarrow \cfrac{E[Y_{11}] - E[Y_{10}]}{E[Y_{10}]}  = \cfrac{E[Y_{01}] - E[Y_{00}]}{E[Y_{00}]}  \nonumber \\
&\Rightarrow \cfrac{\Delta_1 }{E[Y_{10}]}  = \cfrac{\Delta_0}{E[Y_{00}]}   \label{Eq:applogwork1} \\
&\Rightarrow \Delta_1 E[Y_{00}]  = \Delta_0 E[Y_{10}]   \nonumber \\
\text{(minus } \Delta_0 E[Y_{00}] \text{ from both sides)}\qquad  &\Rightarrow (\Delta_1-\Delta_0 ) E[Y_{00}]  = \Delta_0 (E[Y_{10}]-E[Y_{00}])   \nonumber \\
 &\Rightarrow (\Delta_1-\Delta_0 )  = \Delta_0 \cfrac{(E[Y_{10}]-E[Y_{00}])}{E[Y_{00}]}    \label{Eq:applogwork2} \\
 \text{[Express in terms of (additive) DD model parameters]} \qquad &\Rightarrow \alpha_4 = \alpha_3 \cfrac{\alpha_2}{\alpha_1}  \label{Eq:logwork3}
 \end{align}
\noindent{}Combining Equation~(\ref{Eq:applevelwork1}) and Equation~(\ref{Eq:applogwork2}) yields a condition for $sign(\alpha_4) \neq sign(\beta_4)$:\\

\noindent{}when $0 < (\Delta_1-\Delta_0 )  < \Delta_0 \cfrac{(E[Y_{10}]-E[Y_{00}])}{E[Y_{00}]} $ we will have $sign(\alpha_4) \neq sign(\beta_4).$\\

\noindent{}Next, consider the case when the level-DD coefficient is negative i.e.,  $\alpha_4<0$. In this case, there will be discordance in signs when $\beta_4>0$ or $g_T>g_C$, which means the left-hand side of Equation~(\ref{Eq:applogwork2}) is greater than the right-hand side. This leads to a sign-switch condition for the case where $\alpha_4<0$:\\

\noindent{}when $ \Delta_0 \cfrac{(E[Y_{10}]-E[Y_{00}])}{E[Y_{00}]} < (\Delta_1-\Delta_0 )  < 0 $ we will have $sign(\alpha_4) \neq sign(\beta_4).$\\

\noindent{}The two conditions can be summarized simply as the following condition, which completes the proof:\\

\noindent{}when $0 < \left|\Delta_T-\Delta_C\right|  < \left|\Delta_C \cfrac{(E[Y_{T0}]-E[Y_{C0}])}{E[Y_{C0}]}\right| \, ,$ we will have $sign(\alpha_4) \neq sign(\beta_4)$.\\

\clearpage

\beginappendixB
\section{\label{sec:summstats}DD Research Design Literature Review}
\begin{center}
  \begin{table}[htb] \centering
\newcolumntype{C}{>{\centering\arraybackslash}X}

\caption{\label{tab:QJE_2018_2022_1}Articles Using a DD Design Published 2018-2022 in \emph{The Quarterly Journal of Economics}}
{\scriptsize
\begin{tabularx}{\linewidth}{cCCcCC}

\toprule
{\textbf{Article}}&{\textbf{Any Continuous Outcome?}}&{\textbf{Ever Log Outcome?}}&{\textbf{Article}}&{\textbf{Any Continuous Outcome?}}&{\textbf{Ever Log Outcome?}} \tabularnewline
\midrule \addlinespace[\belowrulesep]
\citet{qjx028}&0&--&\citet{qjaa031}&1&1 \tabularnewline
\citet{qjx029}&1&1&\citet{qjaa032}&1&\(0^{\dagger}\) \tabularnewline
\citet{qjx041}&1&0&\citet{qjaa038}&1&1 \tabularnewline
\citet{qjx045}&1&0&\citet{qjaa044}&1&1 \tabularnewline
\citet{qjx040}&1&0&\citet{qjaa040}&1&0 \tabularnewline
\citet{qjx046}&1&0&\citet{qjaa046}&1&0 \tabularnewline
\citet{qjy005}&1&1&\citet{qjab004}&1&0 \tabularnewline
\citet{qjy011}&1&0&\citet{qjab015}&1&1 \tabularnewline
\citet{qjy021}&1&1&\citet{qjab016}&1&1 \tabularnewline
\citet{qjy020}&1&1&\citet{qjab019}&1&1 \tabularnewline
\citet{qjy019}&1&1&\citet{qjab022}&1&1 \tabularnewline
\citet{qjz004}&0&--&\citet{qjab026}&1&1 \tabularnewline
\citet{qjz011}&1&1&\citet{qjab029}&1&1 \tabularnewline
\citet{qjz014}&1&1&\citet{qjab028}&1&1 \tabularnewline
\citet{qjz020}&1&0&\citet{qjab027}&1&1 \tabularnewline
\citet{qjz027}&1&1&\citet{qjab040}&1&1 \tabularnewline
\citet{qjz032}&1&1&\citet{qjab043}&1&0 \tabularnewline
\citet{qjz035}&1&\(0^{\dagger}\)&\citet{qjab045}&1&0 \tabularnewline
\citet{qjz034}&1&1&\citet{qjab049}&1&1 \tabularnewline
\citet{qjz044}&1&1&\citet{qjac006}&1&1 \tabularnewline
\citet{qjaa013}&1&1&\citet{qjac008}&1&1 \tabularnewline
\citet{qjaa014}&1&1&\citet{qjac016}&1&1 \tabularnewline
\citet{qjaa016}&1&0&\citet{qjac019}&1&0 \tabularnewline
\citet{qjaa024}&1&1&\citet{qjac018}&0&-- \tabularnewline
\citet{qjaa027}&1&0&&& \tabularnewline
\bottomrule \addlinespace[\belowrulesep]

\end{tabularx}
\begin{flushleft}
\scriptsize \textbf{Notes}: This table displays summary information on all articles published in \emph{The Quarterly Journal of Economics} over the five year period 2018-2022 that used a difference-in-dffierence research design. Any Continuous Outcome? is 0 if all outcomes are binary, and 1 otherwise. For articles that included at least one continuous outcome, Ever Log Outcome? is zero if the continuous outcome is never presented in log form, and 1 otherwise. \(^{\dagger}\) denotes cases where a continuous outcome is present, and although the outcome is not logged, it is presented in inverse hyperbolic sine form.
\end{flushleft}
}
\end{table}

\end{center}
\clearpage
%

\beginappendixC
\section{\label{sec:summstats}Additional Results}
\begin{center}
  \begin{table}[htb] \centering
\newcolumntype{C}{>{\centering\arraybackslash}X}

\caption{\label{tab:simulations_table_2}Simulating a Zero Levels DD Effect That Yields a Postive or  Negative Log Effect}
{\footnotesize
\begin{tabularx}{\linewidth}{lCCCCC}

\toprule
\multicolumn{1}{c}{ }& \multicolumn{3}{c}{{\textbf{Positive Log Effect}}} &  \multicolumn{2}{c}{\textbf{Negative Log Effect}} \tabularnewline  \cmidrule(l{2pt}r{5pt}){2-4} \cmidrule(l{2pt}r{5pt}){5-6}  \addlinespace[-2ex] \tabularnewline
{}&{(1)}&{(2)}&{(3)}&{(4)}&{(5)} \tabularnewline
\midrule \addlinespace[\belowrulesep]
\textbf{Level-Dependent Variable:}&&&&& \tabularnewline
DD Estimate [\(\hat{\alpha}_4\)]&--0.00&--0.00&0.00&0.00&--0.00 \tabularnewline
&(0.00)&(0.00)&(0.00)&(0.00)&(0.00) \tabularnewline
\textbf{Log-Dependent Variable:}&&&&& \tabularnewline
DD Estimate [\(\hat{\beta}_4\)]&0.154***&0.080***&0.154***&--0.087***&--0.154*** \tabularnewline
&(0.001)&(0.000)&(0.000)&(0.000)&(0.000) \tabularnewline
exp(\(\hat{\beta}_4\))--1&0.167***&0.083***&0.167***&--0.083***&--0.143*** \tabularnewline
&(0.001)&(0.000)&(0.000)&(0.000)&(0.000) \tabularnewline
\addlinespace[.5ex] \midrule \addlinespace[1ex] \(\overline{Y}_{C0}\)&10.00&10.00&20.00&10.00&20.00 \tabularnewline
\(\overline{Y}_{C1}\)&12.00&12.00&24.00&12.00&28.00 \tabularnewline
\addlinespace[.25ex] \(\overline{Y}_{T0}\)&5.00&6.67&10.00&20.00&40.00 \tabularnewline
\(\overline{Y}_{T1}\)&7.00&8.67&14.00&22.00&48.00 \tabularnewline
\((g_T - g_C)/g_C\)&0.167&0.083&0.167&--0.083&--0.143 \tabularnewline
\bottomrule \addlinespace[\belowrulesep]

\end{tabularx}
\begin{flushleft}
\scriptsize \textbf{Notes}: Results based on 10000 simulation runs. DD estimates for both a level- and log-dependent variable are presented in each column. The tables display the mean and (in parentheses) bootrapped standard error of the DD estimates across all simulation runs. At the base of the table, the four elements of the DD are presented for reference. The sample size is 40000 in each simulation. Proportion treated and proportion in the post period are .5 and .5 respectively. The standard deviation of the error term is .2 in all simulations.
\end{flushleft}
}
\end{table}

\end{center}
\clearpage

\beginappendixD
\section{\label{sec:summstats}Summary Statistics}

\subsection{\label{sec:summstatsGR}Male Earnings and The Great Recession}

\subsubsection{Long-Run Trends in Outcomes}

\begin{figure}[H]
  \centering
    \caption{Mean Annual Earnings By Race}
  \includegraphics[width=.9\textwidth]{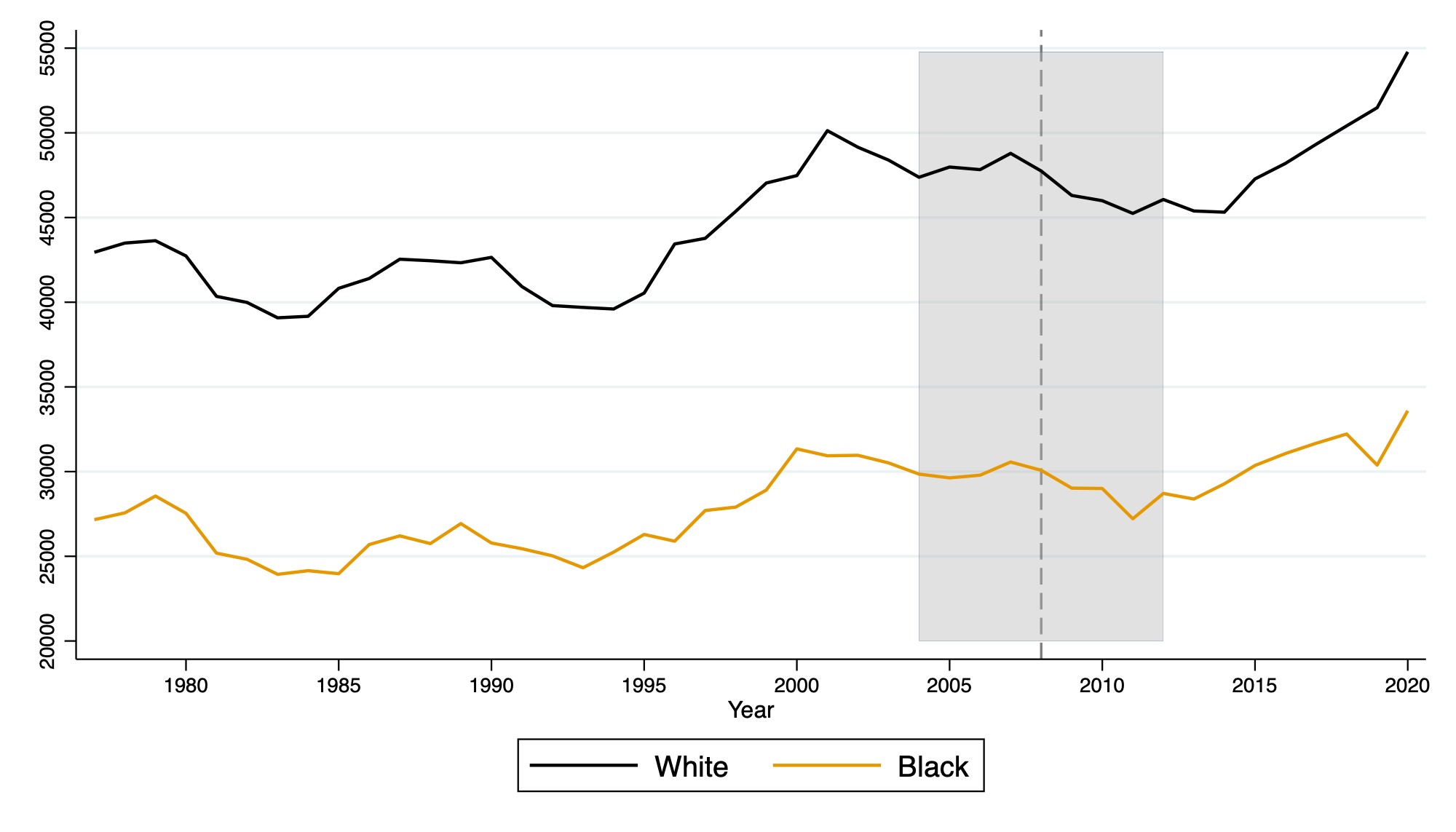}
  \label{fig:GRtrends}
\end{figure}
\clearpage
\subsubsection{Balance Tables}
\begin{center}
  \begin{table}[htbp] \centering
\newcolumntype{C}{>{\centering\arraybackslash}X}

\caption{\label{tab:balance_table_CPS_1}Balance Tests for CPS Data}
{\footnotesize
\begin{tabularx}{\linewidth}{lCCCCCCC}

\toprule
&{(1)}&{(2)}&{(3)}&{(4)}&{(5)}&{(6)}&{(7)} \tabularnewline \midrule
\multicolumn{1}{c}{ }& \multicolumn{3}{c}{\textbf{White}}& \multicolumn{3}{c}{\textbf{Black}}  \tabularnewline  \cmidrule(l{2pt}r{5pt}){2-4} \cmidrule(l{2pt}r{5pt}){5-7}   \addlinespace[-2ex] \tabularnewline
{}&{Pre}&{Post}&{\(p\)-value: Difference}&{Pre}&{Post}&{\(p\)-value: Difference}&{\(p\)-value: DD} \tabularnewline
\midrule \addlinespace[\belowrulesep]
Sample Size&115,525&113,577&&17,846&19,244&& \tabularnewline
\addlinespace[.25ex] Potential Experience&23.2&23.7&[.000]&22.5&23.3&[.000]&[.043] \tabularnewline
&(11.3)&(11.7)&&(11.2)&(11.5)&& \tabularnewline
\addlinespace[.5ex] Education:&&&&&&& \tabularnewline
\, \, \(\leq\) High School&.0757&.0645&[.000]&.147&.129&[.000]&[.106] \tabularnewline
\, \, High School&.318&.311&[.004]&.4&.394&[.297]&[.941] \tabularnewline
\, \, Some College&.269&.274&[.024]&.27&.29&[.000]&[.008] \tabularnewline
\, \, College&.221&.231&[.000]&.13&.128&[.685]&[.010] \tabularnewline
\, \, Postgraduate&.117&.12&[.067]&.053&.0588&[.028]&[.357] \tabularnewline
\addlinespace[.25ex] Married&.65&.632&[.000]&.443&.419&[.000]&[.362] \tabularnewline
\addlinespace[.5ex] Metro Status:&&&&&&& \tabularnewline
\, \, \(\leq\) Non-Metro&.184&.177&[.000]&.106&.0987&[.030]&[.904] \tabularnewline
\, \, Central City&.196&.204&[.000]&.441&.433&[.188]&[.016] \tabularnewline
\, \, Outside Central City&.452&.453&[.818]&.35&.36&[.077]&[.125] \tabularnewline
\bottomrule \addlinespace[\belowrulesep]

\end{tabularx}
\begin{flushleft}
\scriptsize \textbf{Notes}: Means and standard deviations (in parentheses for continuous covariates) are shown. \(p\)-values are based on OLS regressions with Eicker-Huber-White standard errors. Data used: CPS 2005-2012.
\end{flushleft}
}
\end{table}

\end{center}
\subsubsection{Data and Sample Selection}
I use the ASEC March supplement of the CPS for the years 2005-2012, and restrict the  sample to males, aged 25-65, who are either white non-Hispanic or Black. I omit those housed in group quarters. I additional remove those in the army, agricultural workers, those in the private household sector, and the self-employed. In order to maintain the same sample for both level and log specifications, I restrict total income to be strictly positive.
\clearpage
\subsection{\label{sec:trendsEarnings}London House Prices and the Brexit Vote}

\subsubsection{Long-Run Trends in Outcomes}

\begin{figure}[H]
  \centering
    \caption{Mean House Prices in Inner and Outer London -- All Properties}
  \includegraphics[width=.9\textwidth]{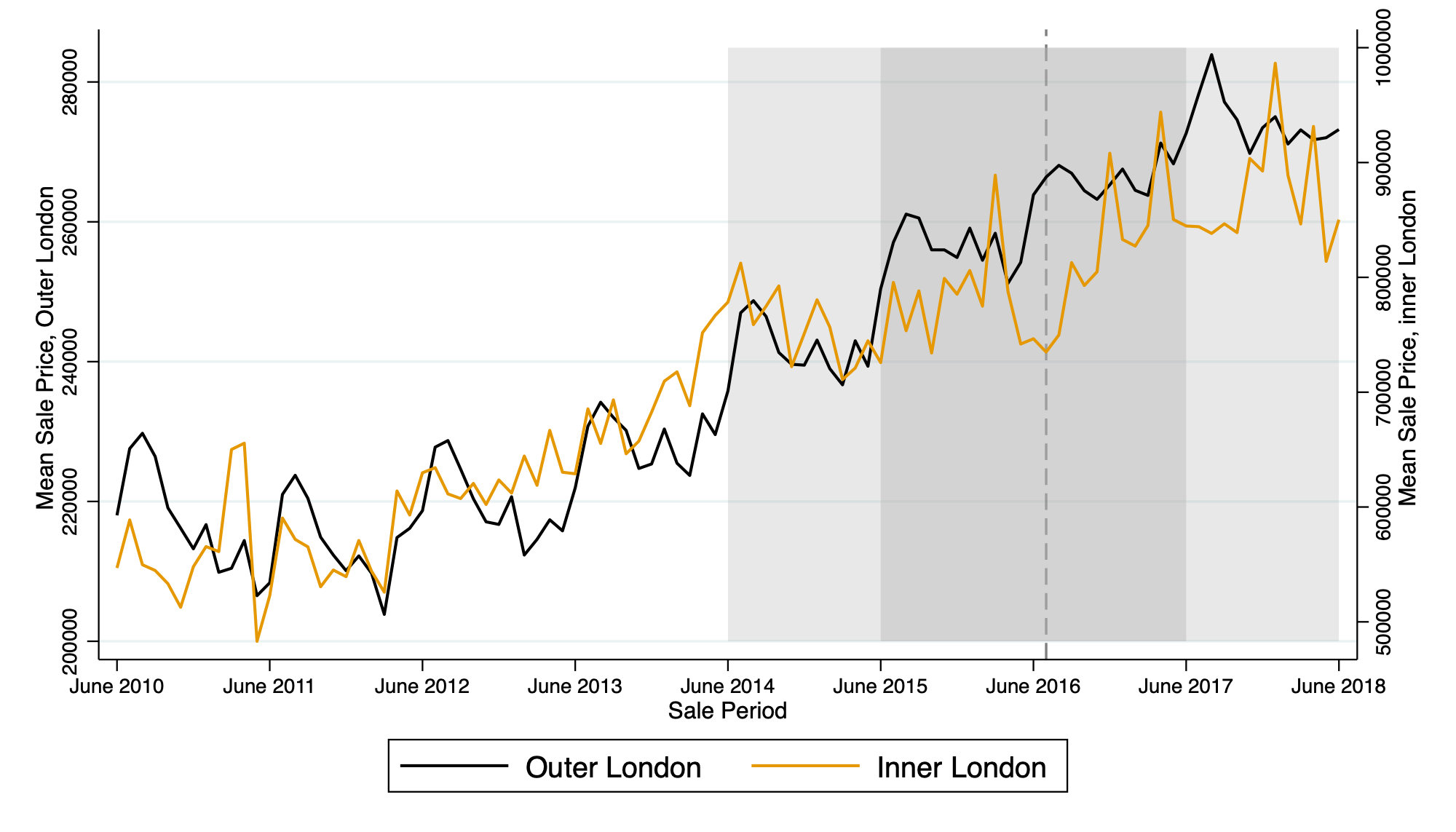}
  \label{fig:BrexittrendsAll}
\end{figure}
\begin{figure}[H]
  \centering
    \caption{Mean House Prices in Inner and Outer London -- Apartments Only}
  \includegraphics[width=.9\textwidth]{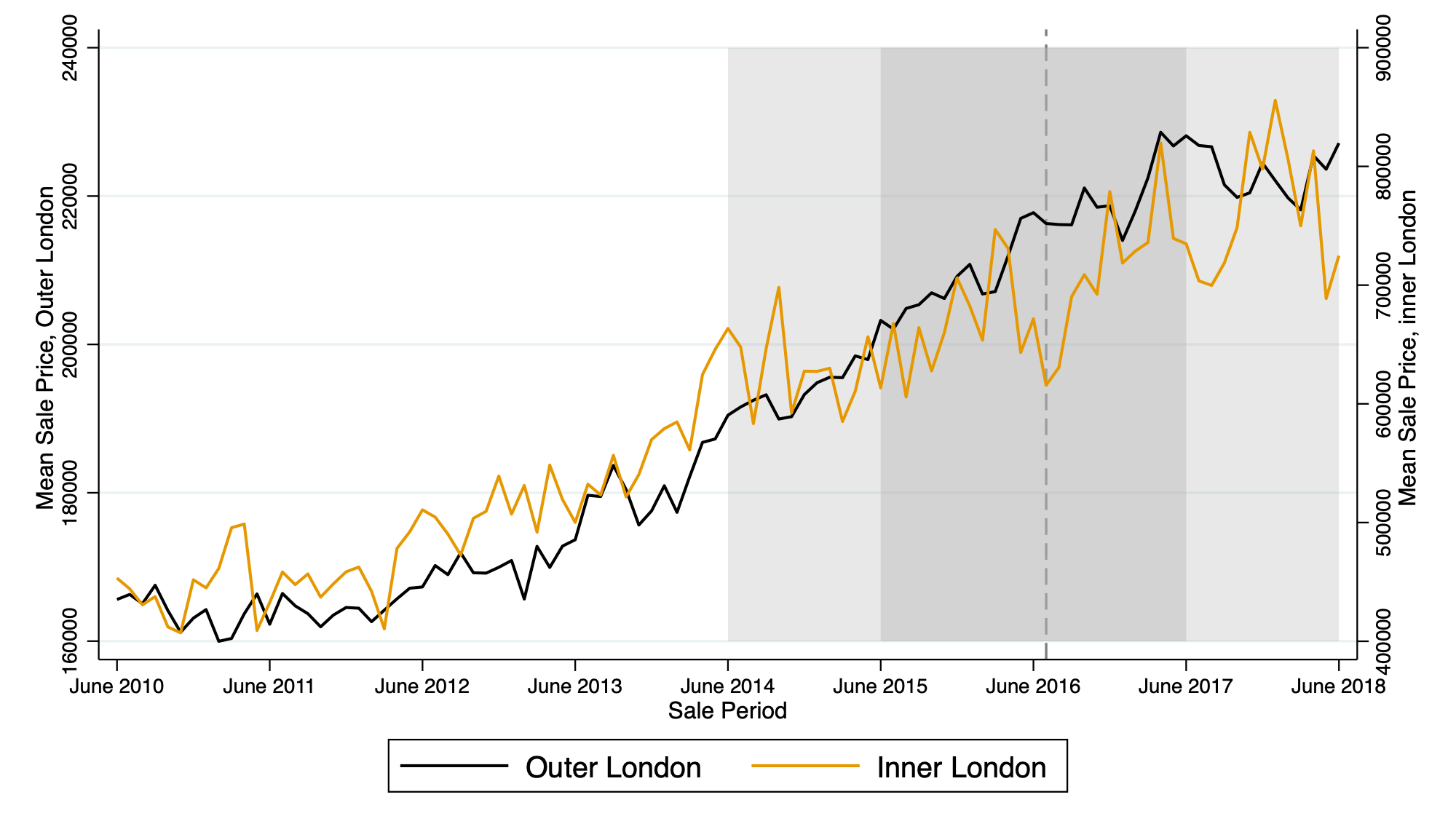}
  \label{fig:BrexittrendsApart}
\end{figure}
\clearpage
\subsubsection{Balance Tables}

\begin{center}
  \begin{table}[htbp] \centering
\newcolumntype{C}{>{\centering\arraybackslash}X}

\caption{\label{tab:balance_table_pricepaid_1}Balance Tests for House Sales Data}
{\footnotesize
\begin{tabularx}{\linewidth}{lCCCCCCC}

\toprule
&{(1)}&{(2)}&{(3)}&{(4)}&{(5)}&{(6)}&{(7)} \tabularnewline \midrule
\multicolumn{1}{c}{ }& \multicolumn{3}{c}{\textbf{Outer London}}& \multicolumn{3}{c}{\textbf{Inner London}}  \tabularnewline  \cmidrule(l{2pt}r{5pt}){2-4} \cmidrule(l{2pt}r{5pt}){5-7}   \addlinespace[-2ex] \tabularnewline
{}&{Pre}&{Post}&{\(p\)-value: Difference}&{Pre}&{Post}&{\(p\)-value: Difference}&{\(p\)-value: DD} \tabularnewline
\midrule \addlinespace[\belowrulesep]
Sample Size&108,268&96,396&&56,386&48,912&& \tabularnewline
\addlinespace[.25ex] New Build&.105&.139&[.000]&.169&.227&[.000]&[.000] \tabularnewline
\addlinespace[.5ex] Property Type:&&&&&&& \tabularnewline
\, \, Detached House&.0662&.0625&[.001]&.00727&.00634&[.065]&[.021] \tabularnewline
\, \, Semi-Detached House&.184&.189&[.004]&.0271&.0281&[.355]&[.044] \tabularnewline
\, \, Townhouse/Terraced&.302&.295&[.000]&.169&.167&[.319]&[.112] \tabularnewline
\, \, Apartments&.448&.454&[.007]&.797&.799&[.354]&[.275] \tabularnewline
\addlinespace[.25ex] Leasehold&.458&.466&[.000]&.806&.808&[.419]&[.078] \tabularnewline
\bottomrule \addlinespace[\belowrulesep]

\end{tabularx}
\begin{flushleft}
\scriptsize \textbf{Notes}: Means are shown. \(p\)-values are based on OLS regressions with Eicker-Huber-White standard errors. Data used: HM Land Registry, December 24, 2014 - December 23, 2017.
\end{flushleft}
}
\end{table}

\end{center}

\subsubsection{Data Sample Selection}
I use the Land Registry Price Paid data for a 2 year window around the Brexit vote, which occurred on 23 June 2016. The Price Paid data covers almost every house sale in England and Wales.\footnote{Source: \url{https://www.gov.uk/government/statistical-data-sets/price-paid-data-downloads}, HM Land Registry.}\footnote{The Land Registry list reasons for the minority of sales that are not registered at \url{https://www.gov.uk/guidance/about-the-price-paid-\#data-excluded-from-price-paid-data}.} 
The data is limited in terms of the household characteristics it contains (indicators for whether the property is a leasehold, if the property is a new-build, as well as property type). In order to compensate for the paucity of household characteristic controls, I specify an extremely low-level spatial fixed effect at the level of Output Area, akin to a census block in the US. Output Areas (OA) are the smallest census-based geographical unit -- there are 181,408 of these in England and Wales, with an average population of 309 at the 2011 census.\footnote{\url{https://www.ons.gov.uk/peoplepopulationandcommunity/populationandmigration/populationestimates/bulletins/2011censuspopulationandhouseholdestimatesforsmallareasinenglandandwales/2012-11-23}} 
.
\clearpage
\end{document}